\documentclass{aastex701}

\begin{document}

\title{A preliminary orbit for the satellite of dwarf planet (136472) Makemake}

\author[orcid=0000-0002-9138-2942]{Daniel Bamberger}
\affiliation{Northolt Branch Observatories, 35039 Marburg, Germany}
\email[show]{danielpeter1204@aol.com}

\begin{abstract}

I present a preliminary orbit for the satellite of dwarf planet (136472) Makemake, based on archival Hubble Space Telescope images taken on 13 days between April 2015 and February 2019. The satellite was detected on twelve of them. A best-fit circular orbit has a period of $18.023 \pm 0.017$ d, a semi-major axis of $22250 \pm 780$ km, and an inclination of $83.7^{\circ} \pm 1.0^{\circ}$ relative to the line of sight. That orbit is nearly edge-on, raising the possibility of ongoing or imminent mutual events between Makemake and its satellite.

\end{abstract}


\section{Introduction} 

The Hubble Space Telescope (HST) has imaged the dwarf planet (136472) Makemake on 13 days between April 2015 and February 2019, using its Wide Field Camera 3 (WFC3). The images not only confirm that Makemake has a satellite, officially designated S/2015 (136472) 1 and henceforth referred to as MK2,\footnote{The unofficial nickname is due to the discovery team, compare hubblesite.org/contents/news-releases/2016/news-2016-18.html.} but also constrain its orbital parameters.

Unfortunately, the data has remained unused since the initial announcement of the discovery of MK2 in April 2016 \citep{Parker16} and the subsequent announcement of its characterization at a conference in 2018 \citep{Parker18}. With the tenth anniversary of the discovery behind us, and no indication that the discoverers still intend to publish the results of their characterization, it is time to raise awareness of the existence of those data, combined with a preliminary analysis. The timing is critical because a predicted series of mutual events between Makemake and MK2 may be approaching, if it is not already ongoing. The nominal orbit suggests that mutual events should occur approximately $7 \pm 2$ years before or after January 2018. The latter would correspond to some time between 2023 and 2027. There may be just enough time to coordinate an observing campaign and, with some luck, catch some of the mutual events.\footnote{Most of this analysis was done in 2018 and early 2019, but was not published, because I wanted to give the observers the opportunity to use the data themselves. I think enough time has passed to publish my own results. Ultimately, it was the timing of the possible mutual events that persuaded me to publish.}

My main focus, apart from the possibility of mutual events, is to illustrate the largely untapped potential of the available data. I do not intend to give a definitive answer about the orbital parameters of MK2 or any conclusions that might be drawn from those. Observations of Makemake with instruments suitable for detecting mutual events are highly encouraged, as is a more in-depth analysis of the data described here. The James Webb Space Telescope could be used to further constrain the orbit, disambiguate its orientation, and determine whether mutual events are observable.

\section{Observations and measurements}

The data consists of sets of exposures of 12×40 seconds and longer exposures (70 and 180 seconds) with different filters. The time between the first and last exposures ranges from about 40 to 120 minutes. I aligned the available 40-second exposures and created a median stack, which I subtracted from each individual exposure to remove the glare from Makemake itself. Then I stacked the twelve processed subs from each set. Of the 13 resulting images, MK2 is visible in twelve of them. In one image, it is too close to Makemake. The longer exposures were discarded here; they will be useful for photometric characterization, which is beyond the scope of this paper. With an orbital period of several weeks, minimal trailing is expected, and I assume MK2 to be stationary in each set.

I measured the offsets of MK2 from Makemake. Each point was then rotated to account for the telescope's orientation and converted to a projected distance between Makemake and MK2, using the known pixel scale (0.04 arc-seconds) and accounting for the distance between Earth and Makemake. This was then fitted to a circular orbit with an assumed measurement precision of 0.7 pixels in both the x- and y-directions.

\begin{table}[ht!]
\begin{tabular}{@{}llllllll@{}}
\toprule
File set & Date (UTC) & Distance (AU) & Angle ($^{\circ}$) & $\Delta$x (px) & $\Delta$y (px) & x' (km) & y' (km) \\
\hline
ICOH03... & 2015-04-27 13:46 & 51.692 & 13.77 & +0.171 & +14.482 & -4920 & +21155 \\
ICOH04... & 2015-04-29 18:17 & 51.712 & 19.77 & - & - & - & - \\
IDMG01... & 2018-03-01 00:50 & 51.715 & 131.42 & -10.472 & +6.781 & +2763 &	-18512 \\
IDMG02... & 2018-03-04 16:12 & 51.694 & 106.81 & +6.036 & +0.441 & -3250 & +8475 \\
IDMG03... & 2018-03-07 18:53 & 51.679 & 121.89 & +14.763 & -5.460 & -4744 & +23117 \\
IDMG04... & 2018-03-11 11:53 & 51.663 & 86.58 & +5.510 & +0.230 & +148 & +8263 \\
IDMG05... & 2018-03-14 12:59 & 51.653 & 85.61 & -9.041 & -3.346 & +3962 & -13893 \\
IDMG06... & 2018-03-18 04:22 & 51.645 & 100.53 & -14.330 & +0.445 & +3268 & -21230 \\
IDMG07... & 2018-03-21 23:11 & 51.640 & 68.31 & +0.975 & +3.213 & -3933 & +3137 \\
IDMG08... & 2018-05-21 03:35 & 52.030 & 357.63 & -2.996 & +12.297 & -3752 & +18732 \\
IDX701... & 2019-01-24 14:03 & 52.121 & 145.60 & +8.817 & -6.578 & -5381 & +15738 \\
IDX702... & 2019-02-02 06:11 & 52.010 & 148.73 & -7.459 & +6.218 & +4748 & -13862 \\
IDX703... & 2019-02-11 07:53 & 51.907 & 113.85 & +9.478 & -0.963 & -4445 & +13639 \\
\end{tabular}
\caption{Measured positions of MK2. The angle is the orientation of HST relative to celestial north. $\Delta$x/$\Delta$y are the offsets from Makemake in the raw images. x'/y' are the offsets after accounting for distance and telescope orientation. Positive x' is west, positive y' is north.}
\end{table}

\begin{figure*}[ht!]
\plotone{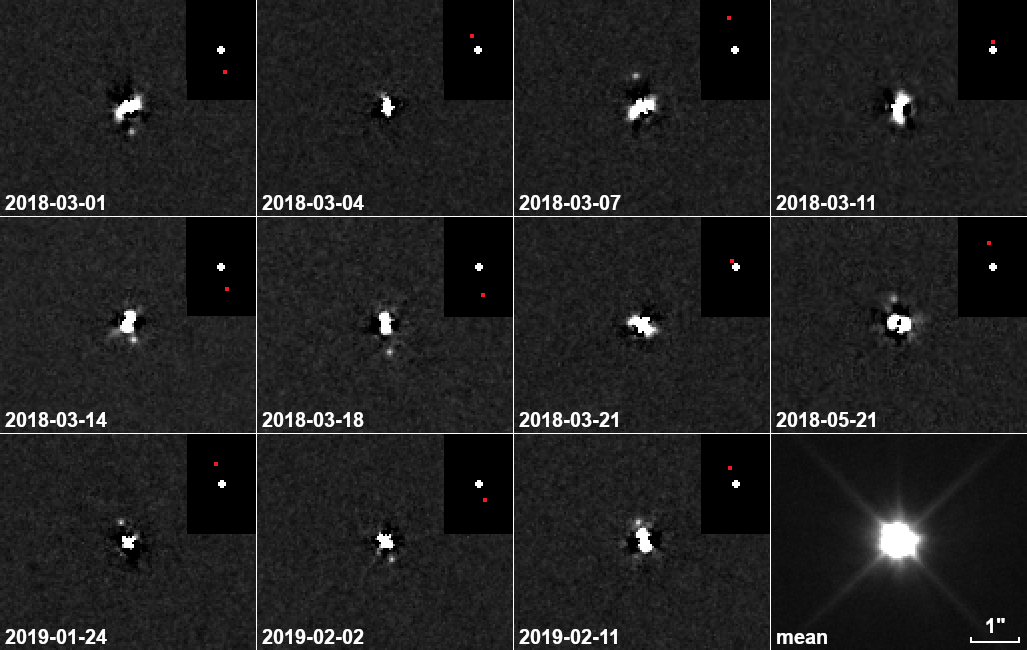}
\caption{Makemake and MK2, imaged in 2018 and 2019. North is up and east is left. The white and red dots in the insets in the upper right corner of each frame indicate the predicted positions of Makemake and MK2, respectively, based on the nominal orbit derived in this paper.}
\end{figure*}

\section{Results}

I find the following best-fit orbital parameters: Orbital period $18.023 \pm 0.017$ d, semi-major axis $22250 \pm 780$ km, inclination $83.7^{\circ} \pm 1.0^{\circ}$ relative to the line of sight (with $90^{\circ}$ corresponding to an orbit that is being viewed edge-on). The orientation of the orbit relative to Makemake (prograde vs. retrograde) is ambiguous.

Assuming a radius of $715 \pm 15$ km for Makemake \citep{Brown13} and that the entire mass of the system is concentrated in the primary, this orbit corresponds to a combined mass of $(2.69 \pm 0.20)\cdot10^{21}$ kg and a density of $1.76 \pm 0.17$ g/cm³.

In March 2018, Parker and colleagues gave a slightly higher density of $2.1$ g/cm³ from their own orbit determination \citep{Parker18}. Using the data that were available at the time, I also find a best-fit density of $2.1 \pm 0.4$ g/cm³. As first noted by them, the nearly edge-on orbit opens the possibility that Makemake and MK2 may undergo a series of mutual events in the near future, or may have done so in the very recent past. Mutual events between Makemake and MK2 would last about 140 minutes at maximum and would occur twice per orbit.

Like the mutual events between Pluto and Charon between 1985 and 1990 \citep{Binzel89}, observations of such events may allow the detection of large-scale surface features and put constraints on the sizes of the two bodies. Given that the latest images are several years old, it is possible that such mutual events are ongoing right now.

\begin{figure*}[ht!]
\plotone{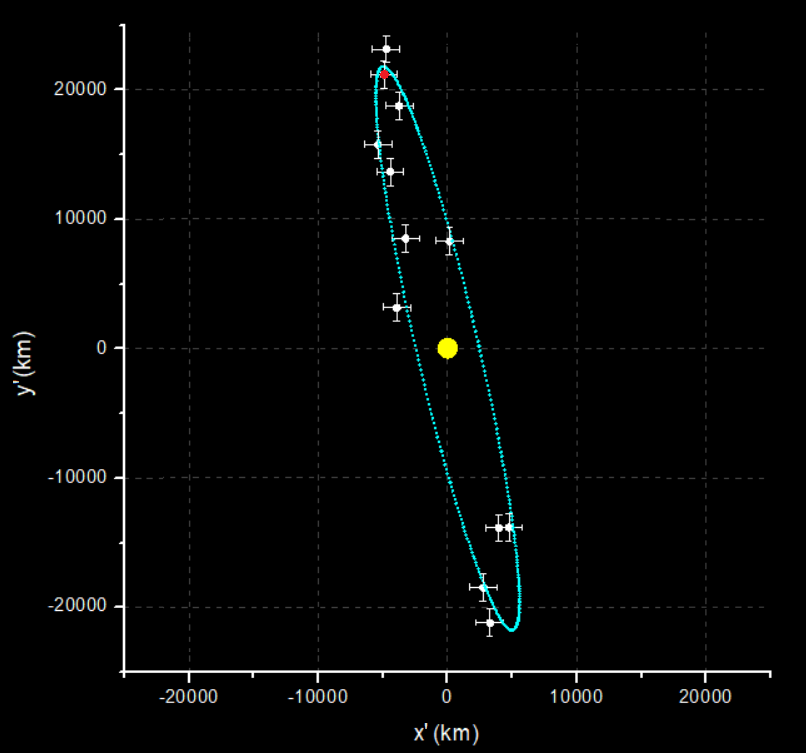}
\caption{The nominal circular orbit of MK2, with the detections that were used for the fit. The red point corresponds to the discovery observation (27 April 2015). North is up and east is left.}
\end{figure*}

\begin{acknowledgments}
I thank the Space Telescope Science Institute (STScI), as well as the principal investigators of the three programs that I used in this paper: Program 13668 by Marc Buie, and programs 15207 and 15500 by Alex H. Parker. All image settings were selected by them. The images were taken by the Hubble Space Telescope and made available at STScI. The raw data as well as STScI's pre-processed images are in the public domain.
\end{acknowledgments}

\facilities{HST(WFC3)}

\bibliography{Makemake_moon_orbit}{}
\bibliographystyle{aasjournalv7}

\end{document}